\documentclass[aps,a4paper,showpacs,preprintnumbers,amsmath,amssymb,floatfix]{revtex4}
\usepackage{dcolumn}
\usepackage{bm}
\usepackage{amsmath,amssymb}
\usepackage{booktabs,subfigure}
\usepackage{graphicx,psfrag}
\usepackage{epsfig}
\usepackage{color} 
\usepackage{rotating}
\usepackage{axodraw}
\usepackage{slashed}
\usepackage{geometry}
\usepackage{rotating}
\usepackage{subfigure}
\allowdisplaybreaks[3]
\newcommand{\wbar}[1]{\mkern1mu\overline{\mkern-3mu#1\mkern-1mu}\mkern-2mu}

\numberwithin{equation}{section} 
\def\lsim{\raise0.3ex\hbox{$\;<$\kern-0.75em\raise-1.1ex\hbox{$\sim\;$}}}
\def\gsim{\raise0.3ex\hbox{$\;>$\kern-0.75em\raise-1.1ex\hbox{$\sim\;$}}}
\newcommand{\beq}{\begin{equation}}
\newcommand{\eeq}{\end{equation}}
\newcommand{\bea}{\begin{eqnarray}}
\newcommand{\eea}{\end{eqnarray}}
\newcommand{\ba}{\begin{array}}
\newcommand{\ea}{\end{array}}
\newcommand{\Neu}[1]{\ensuremath{\widetilde \chi_{#1}^0}}
\newcommand{\Cha}[1]{\ensuremath{\widetilde \chi_{#1}^\pm}}
\begin{document}
\title{ Neutralino and chargino masses and related sum rules beyond MSSM }
\author{Katri Huitu, $^1$
\footnote{\tt{Electronic address: katri.huitu@helsinki.fi}}
P.  N. Pandita, $^2$
\footnote{\tt{Electronic address: ppandita@nehu.ac.in}} and
Paavo Tiitola $^1$
\footnote{\tt{Electronic address: paavo.tiitola@helsinki.fi }}}
\affiliation{ $^1$ Department of Physics, and Helsinki Institute of
Physics,
P. O. Box 64, FIN-00014 University of Helsinki, Finland}
\affiliation{ $^2$ Department of Physics, North Eastern Hill University,
Shillong 793 022, India}

\thispagestyle{myheadings}

\begin{abstract}
\noindent
We study the implications of dimension five operators involving Higgs
chiral superfields for the masses  of neutralinos and charginos in the 
minimal supersymmetric standard model~(MSSM).  These operators can arise
from additional interactions  beyond those of MSSM  involving
new degrees of freedom at or above  the TeV scale. 
In addition to the masses of the neutralinos and charginos, we
study the sum rules involving the masses and squared masses of these 
particles for different gaugino
mass patterns in presence of the dimension five operators.
We derive a relation for the higgsino mixing mass parameter and $\tan\beta$ 
in the presence of the dimension five operators.

\end{abstract}
\maketitle

\section{Introduction}
\label{sec:intro}

Supersymmetry~(SUSY) is a leading candidate~\cite{wess}
for physics beyond the Standard Model~(SM). 
In supersymmetric theories  the Higgs sector, 
so essential for the internal consistency of the SM, is  technically natural~\cite{kaul}.
Supersymmetry is, however,  not an exact symmetry in nature, and the manner in which SUSY is broken is not known.
The necessary SUSY breaking can be introduced through soft
supersymmetry breaking terms that do not 
disturb the stability 
of the hierarchy between the weak scale and the  
large~(grand unified or Planck) scale. The simplest implementation of the idea of low energy broken supersymmetry
is the minimal supersymmetric standard model~(MSSM)
obtained by introducing the supersymmetric partners of the SM states, 
and introducing an additional Higgs doublet with opposite hypercharge to 
that of SM Higgs doublet,
in order to cancel the gauge anomalies and generate masses for 
all the fermions of the Standard Model, with soft supersymmetry breaking 
terms generated by a suitable
supersymmetry breaking mechanism~\cite{Nilles:1983ge}. In order for broken 
supersymmetry to be effective in protecting the weak scale against 
large radiative corrections, the supersymmetric partners of the SM 
particles should have masses  of the order of $\cal{O}$(TeV).  

Because of underlying gauge invariance and supersymmetry, the Higgs sector 
of the MSSM is highly constrained. The LEP lower bound~\cite{LEPhiggsbound} on
the Standard Model Higgs boson is  $m_{h_{\rm SM}} \gsim 114$ GeV.
Although the tree level upper bound $m_h \le M_Z$ on the lightest 
Higgss boson of MSSM is violated by the LEP bound, there are large radiative
corrections to the tree level mass coming from the 
top-stop loops~\cite{oneloophiggs, cponeloop, feynhiggs}. 
If these radiative corrections have to be significant, then
one of the stop mass eigenstates has to be  heavy.  
On the other hand, for these radiative corrections to  
account for the current lower limit on the lightest Higgs boson mass, or for the possible Higgs mass
$m_h\sim 125$ GeV, as hinted by the Large Hadron Collider (LHC) experiments, 
the top squarks must be so massive that it makes the 
MSSM to be finely tuned. Alternatively, there must be large 
left-right mixing  between scalar top quarks.  Such large mixing is 
difficult to obtain in specific models,
and can arise only in special regions in the parameter space \cite{rouven}.
Currently the allowed range of the lightest  Higgs mass 
from the LHC experiments, which is of
interest for  supersymmetric models, extends
from the LEP limit to around 130 GeV \cite{Hlimits_moriond}.
All this suggests that there may be additional degrees of freedom
in the theory beyond those of the MSSM \cite{Dine:2007xi}.
The effect of possible new degrees of freedom, evaluated in terms of 
effective dimension five and six operators have 
been found to be significant for the Higgs boson 
mass, see  {\it e.g.} \cite{Carena:2009gx}. 
As pointed out by Dine, Seiberg and Thomas~\cite{Dine:2007xi},
at dimension five only two operators are relevant for the Higgs 
boson sector.
Several aspects of dimension five operators have been studied in recent 
years, including
neutralino and chargino sector in the context of dark matter
\cite{Berg:2009mq}.

It is interesting to note that there  are several candidates for such 
additional physics beyond 
the MSSM~\cite{nmssm, Higgs1,  Huitu:1997rr, Espinosa:1998re,
Higgs2}. If this new
physics lies at an energy scale  which is above the masses of the
MSSM degrees of freedom~(we call it $M$), then it is convenient to
study the effects of such additional
degrees of freedom by using an effective Lagrangian approach
from which the
physics at scale $M$ has been integrated out. 
The most general superpotential for the MSSM, which involves only the
Higgs chiral superfields, up to dimension
five can be written as~\cite{Dine:2007xi}
\bea
W_5 & = &   \mu H_u H_d + { \lambda \over M} (H_u H_d)^2,
\label{dimen5}
\eea
where $\mu$ is the higgs(ino) mixing parameter in the superpotential of
MSSM,  $M$ is  an energy scale which is much above the typical masses of the
superparticles of MSSM, and $\lambda$ is  a dimensionless coupling. 
It has been shown that the dimension five operator
in (\ref{dimen5}) raises the lightest Higgs boson mass of MSSM
above the LEP limit without fine tuning, and, hence, without loss of 
naturalness~\cite{Dine:2007xi}.


Apart from the supersymmetry conserving dimension five operator
in (\ref{dimen5}), there is another dimension five 
operator which involves supersymmetry breaking and can be represented
by a dimensionless chiral spurion superfield~\cite{Dine:2007xi}. 
However, if  $m_{\rm SUSY} \approx |\mu|$,
the correction to the lightest Higgs mass comes dominantly
from the supersymmetric operator (\ref{dimen5}), thus we will consider the effects of this operator only.

We will assume here that the R-parity, $R_P=(-1)^{3B-2L+2s}$, is conserved,
leading to a stable lightest supersymmetric particle (LSP).
In the models that we will consider, it is the lightest neutralino, and
thus the other R-odd particles will finally decay to it.
Here we will study the  effects of the dimension five operator 
(\ref{dimen5}) on the spectrum of neutralinos and charginos.
In this work we will  concentrate on 
different supersymmetry breaking mechanisms, which lead
to different mass patterns for the gaugino mass parameters, and the 
implications of the  dimension five operator (\ref{dimen5})
for these mass patterns.  In particular, 
we will demonstrate that using sum rules specific for the neutralino and 
chargino sector, one could distinguish  between different breaking patterns
in presence of the dimension 5 operator.
We will also derive a formula for  the $\mu$-parameter as a function 
of $\tan\beta$, and we will also consider determining the amount of 
the dimension five contribution using the sum rules.

In Section~\ref{sec:mass_matrices} we write down
the mass matrices for the neutralinos and charginos in the presence of
the dimension five operator (\ref{dimen5}). We review the experimental 
constraints on the parameters of the neutralino and
chargino mass matrices, and discuss relevant aspects of different patterns 
for the soft supersymmetry
breaking gaugino mass parameters that arise in models of low energy
supersymmetry.
In Section III we present our results for the spectrum of charginos and 
neutralinos, and the effect of dimension five operator on this spectrum. 
Further, we  discuss  sum rules involving the  masses and squared masses
of neutralinos and charginos which can be used to study the effect of the
dimension five operator.  We conclude with a summary in Section IV.

\section{Neutralino and Chargino Mass Matrices}
\label{sec:mass_matrices}

\subsection{Higgsino sector} 
The superpotential (\ref{dimen5}) leads, up to fimension five,
to the following interaction Lagrangian involving only the 
higgsino~($\widetilde{H}_u, \widetilde{H}_d$) and the 
Higgs~($H_u, H_d$) fields~\cite{Dine:2007xi}: 
\bea
{\cal L} & = & \mu (\widetilde{H}_u \widetilde{H}_d)
       - { \epsilon_1 \over \mu*} 
       \left[ 2 (H_u H_d) (\widetilde{H}_u \widetilde{H}_d)
       +2 (\widetilde{H}_u {H}_d ) (H_u \widetilde{H}_d   )
       + (H_u \widetilde{H}_d ) (H_u \widetilde{H}_d )
       + (\widetilde{H}_u H_d) (\widetilde{H}_u H_d )  \right]  ~+~{\rm H.c.},
\nonumber \\
\label{higgsinohiggs}
\eea
where $SU(2)_L$ contraction between the fields in
round parentheses is implied, and where 
\bea
\epsilon_1 = \lambda \mu^*/ M.
\label{eps1}
\eea 
For definiteness, we 
shall take $\mu$ to be real in this paper.

The first and second terms in
(\ref{higgsinohiggs}) with scalar Higgs expectation values
modify the charged and neutral higgsino Dirac masses.
The third and fourth terms in
(\ref{higgsinohiggs}) with scalar Higgs expectation values give
rise to neutral higgsino Majorana masses which are absent
in the tree-level neutralino mass matrix.
Precision fits to both masses and couplings of neutralinos and
charginos would be sensitive to the dimension five Higgs-higgsino
interactions.
It is important to note that the interactions
(\ref{higgsinohiggs}) are all proportional to a single coupling,
$\epsilon_1$, which is the same as the coupling affecting the
Higgs mass~\cite{Dine:2007xi}.

After the electroweak symmetry is broken,
the neutralino mass matrix in the bino-wino-higgsino basis 
following from (\ref{higgsinohiggs}) can be written as~\cite{ Berg:2009mq}
\bea {\mathcal M_0}=\left(
\begin{array}{cccc} M_1 & 0 & -M_Z\cos\beta\sin\theta_W &
M_Z\sin\beta\sin\theta_W\\[2mm]
0 & M_2 & M_Z\cos\beta\cos\theta_W & -M_Z\sin\beta\cos\theta_W\\[2mm]
-M_Z\cos\beta\sin\theta_W& M_Z\cos\beta\cos\theta_W 
& \frac{2\lambda}{M} v^2 \sin^2\beta
 & -\mu + \frac{4\lambda}{M} v^2 \sin\beta \cos\beta   \\[2mm]
M_Z\sin\beta\sin\theta_W & -M_Z\sin\beta\cos\theta_W & 
-\mu + \frac{4\lambda}{M} v^2 \sin\beta \cos\beta  &  
\frac{2\lambda}{M} v^2 \cos^2\beta\\[2mm]
\end{array} \right), \label{neutmatrix}
\eea
where $M_2$ and $M_1$ are the $SU(2)_L$ and $U(1)_Y$ soft supersymmetry
breaking gaugino masses, $M_Z^2 = {1 \over 2} (g^2 + g^{\prime 2}) v^2$,
$M_W^2 = {1 \over 2} g^2 v^2$, $g$ and $g'$ are $SU(2)_L$ and $U(1)_Y$
gauge couplings, and  $v = (2^{3/2} G_F)^{-1/2} \simeq 174$ GeV is the Higgs
vacuum expectation value.  We shall denote the eigenstates of the
neutralino mass matrix by $ \tilde \chi^0_1, \tilde \chi^0_2, \tilde
\chi^0_3, \tilde \chi^0_4$ with eigenvalues $M_{\tilde \chi^{0}_{i =
    1, 2, 3, 4}}$, labeled in order of increasing mass. 

In the wino-higgsino basis, the chargino mass matrix at dimension five can be written as
\bea
{\mathcal
M_\pm} = \left(\begin{array}{cc} M_2 & {\sqrt 2} M_W \sin\beta\\[2mm]
{\sqrt 2} M_W \cos\beta & \mu - \frac{2\lambda}{M} v^2 \sin\beta \cos\beta\\
\end{array} \right). \label{chargematrix}
\eea
We shall denote the eigenstates of the chargino mass
matrix~(\ref{chargematrix}) as
$\tilde \chi^{\pm}_1$ and $\tilde \chi^{\pm}_2$,  with
eigenvalues $M_{\tilde \chi^{\pm}_{i = 1, 2}},$ respectively.

Bounds on $\epsilon_1$ have been discussed in~\cite{Berg:2009mq}.
 The dimension five operator (\ref{dimen5}) causes a shift in the mass of
the lightest Higgs boson,  and if one assumes shift in $m_{h^0}$ to be at most
$20\% - 30\%$,  then $\epsilon_1$ is constrained 
to values smaller than $0.05$~\cite{Berg:2009mq}. 
Larger shifts in the Higgs mass could in principle
disrupt the vacuum stability by creating a new global 
minimum for  the potential. This issue was examined in~\cite{Blum}, 
and a criterion was found to exclude transitions to such a vacuum.
Furthermore, $\epsilon_1$ is restricted  by the scale of new physics 
appearing beyond the MSSM. 
If the scale of new physics is taken to be $M/\lambda>5$ TeV, 
then using (\ref{eps1}) one arrives 
at a limit $|\epsilon_1 |\lsim 0.04$ for $\mu=200$ GeV, whereas $M/\lambda>2$ TeV allows for $|\epsilon_1 |\lsim 0.1$.
This limit is further increased at larger values of $\mu$. 
However, for large $\mu$ the lightest neutralino and chargino are mostly gauginos and the contribution from 
$\epsilon_1$ to their masses is much less significant. 
In the following we limit our discussion to $|\epsilon_1|\lsim 0.1$.

We note that the dimension five operator (\ref{dimen5}) contributes to  
the lower right $2 \times 2$ submatrix of the neutralino mass matrix
(\ref{neutmatrix}). 
Furthermore, this operator also contributes to the $(2,2)$ element of the
chargino mass matrix.
We have included the most significant MSSM one-loop radiative
corrections to the neutralino and chargino mass matrices in our analysis. 
Although these loop corrections
are small~(of the order of few GeV), these  corrections in the
$(3,3)$ and $(4,4)$ elements of the neutralino mass matrix can be
important, since these elements vanish in the absence of dimension 
five contribution. 

\bigskip

\subsection{Experimental Constraints}

Collider experiments have searched for the supersymmetric partners of
the Standard Model particles. No supersymmetric partners of the SM
particles have been found in these experiments.  At present only lower
limits on their masses have have been obtained.  In particular, the
search for the lightest chargino state at LEP has yielded lower
limits on its mass \cite{lep-chargino}.
The limit depends on the spectrum of the model \cite{Yao:2006px}.
Assuming that $m_0$ is large, from the chargino pair production
one obtains the lower bound
\beq 
M_{\tilde \chi_1^{\pm}} \gsim 103~~{\rm GeV}.  \label{ch-limit}
\eeq
For small $m_0$, the bound is lowered, so that for
$m_{\tilde\nu}<200$ GeV, but $m_{\tilde\nu}>m_{\tilde\chi_1^\pm}$,
the limit becomes \cite{Yao:2006px}
\beq M_{\tilde \chi_1^{\pm}} \gsim 85~~{\rm GeV}.  \eeq
For the parameters of the chargino mass matrix (\ref{ch-limit}) implies
an approximative lower limit~\cite{Abdallah:2003xe,Dreiner:2009ic}
\beq M_2,~~ \mu \gsim 100~~{\rm GeV}.
\label{limits1}
\eeq
The limits (\ref{limits1}) on the parameters $M_2$ and $\mu$ are
found from scanning over the MSSM parameter space and are thus model
independent.

Another important constraint for parameters in the SUSY models 
comes from the  mass of the lightest Higgs boson. 
The current lower limit on the mass of the lightest Higgs boson 
from LEP is 114.4 GeV.
Including theoretical uncertainties from NNLO and higher 
corrections \cite{Degrassi:2002fi} 
will decrease the limit by around 3 GeV, and in our calculations 
we will use the lower limit of 111 GeV.
The LHC experiments have found indications for a particle with $m\sim 125$ GeV.
Since this needs to be confirmed, we do not impose this mass
constraint, but we will discuss 
the case of such a Higgs boson.

The LHC experiments have obtained  constraints on the 
the squark and gluino masses. 
The ATLAS and CMS preliminary results indicate \cite{hcp2011}
that in the gravity mediated breaking 
the gluino mass limit is close to 1 TeV for a number of channels.
Since this limit is model dependent, in the plots we will show the ranges for
gaugino mass parameters satisfying Eq.~(\ref{limits1}) but keep in mind that the
small gaugino mass parameters may violate the experimentally measured gluino
mass.

\subsection{ Gaugino Mass Patterns}
Having constrained the parameters $M_2$ and $\mu$, which enter the
chargino as well as the neutralino mass matrix, we now turn to the
theoretical models for the supersymmetry breaking gaugino mass
parameters $M_1, M_2$, and $M_3$. Theoretically, a simple set of
patterns has emerged for these SUSY breaking parameters, which can be
described as follows.
Here we will briefly list the mass patterns. A
more detailed discussion can be found {\it e.g.} in \cite{Choi:2007ka,Huitu:2010me}.

\subsubsection{Gravity mediated breaking}
The first pattern, which has been the object of extensive studies, is the one
which arises in the gravity mediated supersymmetry breaking models, 
usually referred to as the mSUGRA pattern.
In the gravity mediated minimal supersymmetric standard model, the
soft gaugino masses $M_i$ and the gauge couplings $g_i$ satisfy the renormalization group
equations~(RGEs)~($|M_3| \equiv  M_{\tilde g}$, the tree level gluino mass) 
\bea
16\pi^2\frac{dM_i}{dt} & = & 2 b_i M_i g_i^2, ~~~~b_i =
\left(\frac{33}{5}, 1, -3\right), \label{gaugino1}\\
16\pi^2\frac{dg_i}{dt} & = & b_i g_i^3 \label{gauge1}
\eea
at the leading order, where $ i = 1, 2, 3 $ refer to the 
$U(1)_Y, SU(2)_L$ and the $SU(3)$ gauge groups, respectively. 
Furthermore, $g_1 =\frac{5}{3}g',\; g_2 = g$, and $g_3$ 
is the $SU(3)_C$ gauge coupling. 
 With the boundary 
conditions~($\alpha_i = g_i^2/4\pi, \, i = 1, 2, 3$)
\bea
M_1 & = &  M_2 = M_3 = m_{1/2}, \label{gauginogut}\\
\alpha_1 & = & \alpha_2 =  \alpha_3 = \alpha_G \label{gaugegut}
\eea
at the GUT scale $M_G$,  the  RGEs (\ref{gaugino1}) and (\ref{gauge1}) 
imply that the soft supersymmetry breaking gaugino masses scale like
gauge couplings:
\bea
\frac{M_1(M_Z)}{\alpha_1(M_Z)} & = & \frac{M_2(M_Z)}{\alpha_2(M_Z)}
= \frac{M_3(M_Z)}{\alpha_3(M_Z)} =  \frac{m_{1/2}} {\alpha_G}.
\label{gaugino2}
\eea
%
After including radiative corrections, the ratios for gaugino masses are
\begin{equation}
M_1 : M_2 : M_3 \simeq 1 : 1.9 : 6.2.
\label{msugra1}
\end{equation}
This pattern is typical of any scheme obeying 
Eqs. (\ref{gaugino1}) and (\ref{gauginogut}).
Note that the gluino mass used above is the 
running mass evaluated at the scale of the  gluino mass, whereas
the gaugino mass parameters $M_1$ and $M_2$ are running 
parameters evaluated at the weak scale  $M_Z$. 
Using  the  ratio (\ref{msugra1}) and the lower limit (\ref{limits1}), we have
the constraint  
\bea
M_1 & \gsim & 50~ {\rm GeV}, 
\label{msugra2}
\eea 
in the gravity mediated supersymmetry breaking models.  

We note that in the gravity mediated supersymmetry breaking models,
the parameter $\mu$ is not constrained. As such $|\mu |$ can be smaller or
larger than $M_{1,2}$. If $|\mu |\gg M_1, M_2$, then the lightest
neutralino is mostly a gaugino, whereas in the opposite case
 $|\mu |\ll M_1, M_2$, it is dominantly a higgsino.
 
\subsubsection{Anomaly mediated breaking}
A second pattern of gaugino masses, which is distinct from
the mSUGRA pattern, arises in anomaly mediated supersymmetry breaking
models~(AMSB).
Since the soft supersymmetry breaking parameters are
determined by the breaking of the scale invariance, they can be
written in terms of the beta functions and anomalous dimensions in the
form of relations which hold at all energies.  In MSSM, the pure anomaly mediated
contributions to the supersymmetry breaking gaugino masses can be written as~\cite{amsb}
\bea
M_\lambda &=& \frac{\beta_g}{g} m_{3/2},\label{gmass}
\eea
where $m_{3/2}$ is the gravitino mass, 
$\beta$'s are the relevant $\beta$ functions.
We note that the gaugino masses are proportional to their corresponding
gauge group $\beta$ functions with the lightest supersymmetric 
particle being mainly a wino.

However, it turns out that the pure scalar mass-squared anomaly
contribution for sleptons is negative~\cite{RS}.
A simple way to cure the tachyonic spectrum is to add a common mass parameter $m_0$
to all the squared scalar masses \cite{Gherghetta:1999sw}, assuming that such an
addition does not reintroduce the supersymmetric flavor problem. 

In AMSB,  after including radiative corrections,
we have the following pattern for the
gaugino masses: 
\begin{equation}
M_1 : M_2 : |M_3| \simeq 2.8 : 1 : 7.1, 
\label{anomaly1}
\end{equation}
in the minimal supersymmetric standard model with anomaly mediated
supersymmetry breaking.

Using (\ref{limits1}) and the anomaly pattern of the gaugino masses
(\ref{anomaly1}), we have
\bea
M_1 & \gsim & 280~ {\rm GeV}. 
\eea
This is to be contrasted with the corresponding result (\ref{msugra2})
for the gravity mediated supersymmetry breaking. We further note that 
in the anomaly mediated supersymmetry breaking mechanism, the higgs(ino)
parameter $\mu$ cannot be smaller than $M_1$ due to the constraints following
from electroweak symmetry breaking condition~\cite{Gherghetta:1999sw}. 
This implies that the dominant component of the lightest neutralino
will be a gaugino. Thus, the effect of the dimension five operator
on the lightest neutralino mass will be small, since it affects the
higgsino component only.

\subsubsection{Mirage mediated supersymmetry breaking}
A third simple gaugino mass pattern arises from the
mirage (or mixed modulus) mediated  supersymmetry breaking, which is a
hybrid between anomaly mediated supersymmetry breaking and mSUGRA
pattern.
Mirage mediation is naturally realized in KKLT-type
moduli stabilization~\cite{Kachru:2003aw} and its generalizations, 
a well known example
being KKLT moduli stabilization in type IIB string
theory~\cite{Kachru:2002he}.  
Phenomenology and cosmology of mirage
mediation have been studied in \cite{Choi:2006im,
Choi:2005uz,endo05,falkowski05,baer06,baer,yama, kitano-lhc,
kawagoe}. Signatures of this scenario at LHC and the spectrum of
neutralino mass in particular have been studied in~\cite{Choi:2007ka, cho3}.
The boundary conditions for the soft supersymmetry breaking gaugino mass terms can be written as~\cite{choi1}
\begin{eqnarray}
  M_a&=& M_0 \Big[\,1+\frac{\ln({\wbar M}_{Pl}/m_{3/2})}{16\pi^2} b_a
  g_a^2\alpha\,\Big],
  \label{eq:bc1}
\end{eqnarray}
where $M_0\sim 1$ TeV is a mass parameter characterizing the moduli
mediation, ${\wbar M}_{Pl}$ is the reduced Planck mass, $g_a$
are the gauge couplings and $b_{a}$ the corresponding one-loop 
beta function coefficients, and
$\alpha={m_{3/2}}/[{M_0\ln({\wbar M}_{Pl}/m_{3/2})}]={\cal O}(1)$ is a
parameter representing the ratio of anomaly mediation to moduli
mediation. In addition to $M_0$, $\alpha$ and $\tan\beta$, mirage mediation
is parametrized by $a_i$, and $c_i$, for which we follow definitions of \cite{choi1}.

Throughout the paper we have used the values $c_{i}=a_{i}=1$. 
At low energies, the gaugino masses in mirage mediation can be written
as
\begin{equation}
 \frac{M_a(\mu)}{g_a^2(\mu)}\,=\,\left(1+\frac{\ln({\wbar M}_{Pl}/m_{3/2})}{16\pi^2}
    g_{GUT}^2b_a\alpha\right)\frac{M_{0}}{g_{GUT}^2}.
\label{miragelowe}
\end{equation}
This leads to a unification of the soft gaugino masses at the mirage messenger 
scale \cite{mirage2} 
\bea
M_{\rm mir}=M_{GUT}\left(\frac{m_{3/2}}{{\wbar M}_{Pl}}\right)^{\alpha/2},
\label{mirageuni}
\eea 
which is lower than GUT scale for positive values of $\alpha$. 
For $g_{GUT}^2\simeq 1/2$ the resulting low energy values
yield the mirage mass pattern
\begin{equation}
M_1 : M_2 : M_3 \simeq (1 + 0.66 \alpha): (2 + 0.2 \alpha): (6 - 1.8 \alpha).
\end{equation}
Including the radiative corrections  for the gaugino masses, we obtain
\bea
M_1 : M_2 : M_3 \simeq 1 : 1.5 : 2.1 ~~~~{\rm for}~ \alpha = 1, \\
M_1 : M_2 : M_3 \simeq 1 : 1.2 : 0.92 ~~~~{\rm for}~ \alpha = 2.
\label{mirage1}
\eea
where we have used the value $M_0=1$ TeV. Thus, for the mirage mediation, 
we find
\bea
& M_1 \gsim & 67 \; {\rm GeV} ~~~~{\rm for}~ \alpha = 1, \\
& M_1 \gsim & 83 \; {\rm GeV} ~~~~{\rm for}~ \alpha = 2.
\eea
Depending on the values of parameter, the lightest neutralino can be dominantly 
either a higgsino or a gaugino.

\section{Numerical results and sum rules for neutralino and chargino masses}

For large values of $\mu$, the lightest neutralino and chargino are almost 
pure gauginos.
In this case, the corrections to the lightest neutralino and chargino masses 
from BMSSM 
operators are small, since they affect the higgsino sector.
If, on the other hand, the $\mu$ parameter is small compared to the gaugino mass
parameters, {\it i.e.} if the lightest neutralino and chargino are dominantly higgsinos, the BMSSM
corrections to their masses  can be significant. 
 In the case when the lightest neutralino and chargino are dominantly 
gauginos, it may   be possible 
to study the effects of dimension five operator 
by using the sum rules for the masses of all the neutralinos and charginos.
We will demonstrate that sum rules involving the neutralino and chargino masses
can be used to distinguish between
the different SUSY breaking patterns in presence of dimension five operator.

\begin{figure}
  \psfrag{N}{$m_{\Neu{1}}$(GeV)}
  \psfrag{M}{$M_{1}$(GeV)}
  \psfrag{-}{-}
  \psfrag{0}{0}
  \psfrag{1}{1}
  \psfrag{2}{2}
  \psfrag{3}{3}
  \psfrag{4}{4}
  \psfrag{5}{5}
  \psfrag{6}{6}
  \psfrag{7}{7}
  \psfrag{8}{8}
  \psfrag{9}{9}
  \psfrag{mnsugramu200}{}
  \includegraphics[height=5.5cm]{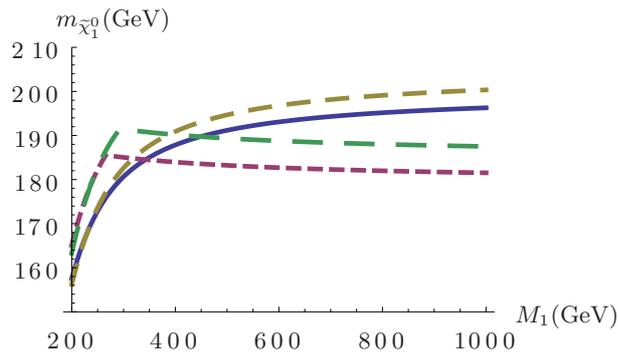}
  
\caption{The lightest neutralino mass in mSUGRA at tree level,
and with one-loop radiative corrections as a function of the gaugino mass
parameter  $M_1$.  The blue solid line corresponds to the tree level mass 
with $\epsilon_1 = 0$. The other curves are in order of increasing dash 
length:  tree level mass with $\epsilon_1 = 0.1$ (violet); one-loop mass
with $\epsilon_1=0$ (ochre); and one-loop mass with $\epsilon_1 = 0.1$ 
(green).  Here $\mu=200$ GeV and tan $\beta=10$.}
\label{fig:radcorr}
\end{figure}

\begin{figure}
  \psfrag{N}{$m_{\Neu{1}}$(GeV)}
  \psfrag{C}{$m_{\Cha{1}}$(GeV)}
  \psfrag{M}{$M_{1}$(GeV)}
  \psfrag{-}{-}
  \psfrag{0}{0}
  \psfrag{1}{1}
  \psfrag{2}{2}
  \psfrag{3}{3}
  \psfrag{4}{4}
  \psfrag{5}{5}
  \psfrag{6}{6}
  \psfrag{7}{7}
  \psfrag{8}{8}
  \psfrag{9}{9}
    
   \subfigure[$\tan\beta=10$]{\includegraphics[height=5.0cm]{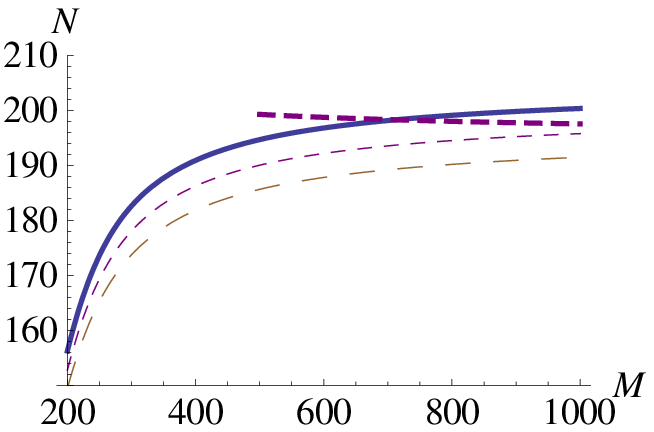}}
   \subfigure[$\tan\beta=30$]{\includegraphics[height=5.0cm]{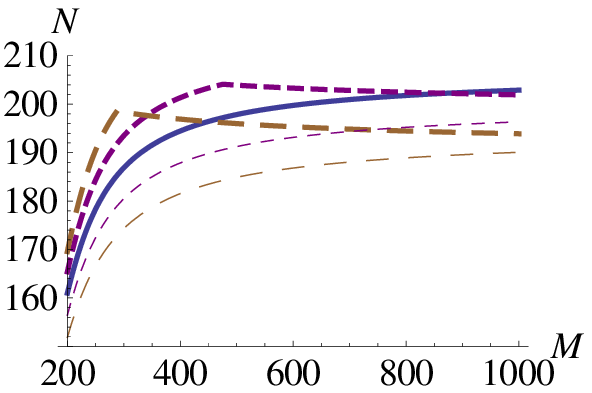}}
   \subfigure[$\tan\beta=10$]{\includegraphics[height=5.0cm]{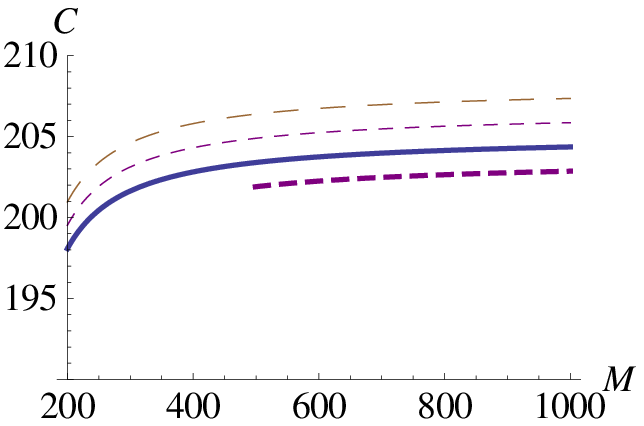}}
   \subfigure[$\tan\beta=30$]{\includegraphics[height=5.0cm]{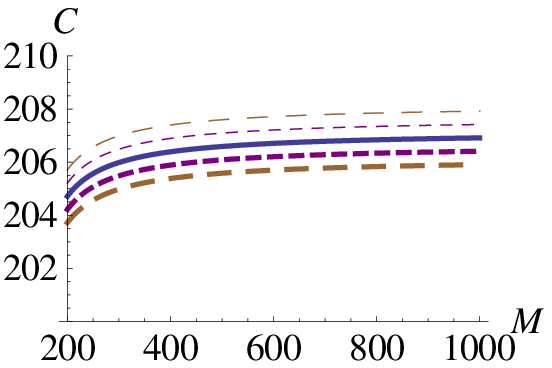}}
 \caption{The lightest neutralino and chargino masses in mSUGRA for several values for the parameter 
$\epsilon_1 = \frac{\lambda}{M}\mu$. The blue solid line corresponds to 
$\epsilon_1 = 0$, and the thick dashed lines in  order of increasing 
dash length represent $\epsilon_1 = 0.05$ (violet), $\epsilon_1 = 0.1$ (ochre). 
The thin dashed lines denote the lightest neutralino mass for  $\epsilon_1 = -0.05$ (violet), $\epsilon_1 = -0.1$ 
(ochre), again in the order of increasing dash length. Here $\mu=200$ GeV and one-loop radiative 
corrections are included.}
\label{fig:mn1}
\end{figure}

Since radiative corrections will be competing with the corrections coming
from the dimension five operators, it is important to compare the magnitude 
of the
$\epsilon_1$ corrections with one-loop radiative corrections.
In Fig.~\ref{fig:radcorr} we have plotted the lightest neutralino mass 
in the mSUGRA pattern of gaugino masses
with $\mu=200$ GeV, $\tan\beta= 10$.
We have plotted the lightest neutralino mass at the tree level, with radiative 
corrections, with corrections coming only  from $\epsilon_1$, and with 
both the radiative and $\epsilon_1$ corrections. 
The radiative corrections are calculated 
using small $\mu$ approximation~\cite{dreesetal, pierceetal}.
Only the contributions
from quark-squark loops are included and squark masses are taken to be
$1$ TeV. 
It is seen that radiative corrections and $\epsilon_1$ corrections are both generally a few GeV,
but for large gaugino mass parameters they are of the opposite sign.
At $M_1=1$ TeV, the radiative and $\epsilon_1$ corrections are of similar magnitude (but opposite sign) for
$\epsilon_1=$-0.04. 
The kink in the BMSSM corrections shows that at the corresponding
value of the  parameter $M_1$, the lightest neutralino changes 
from an  eigenstate containing 
a significant gaugino component to another mass eigenstate, which is almost 
a pure higgsino.

\begin{figure}
  \psfrag{N}{$m_{\Cha{1}}-m_{\Neu{1}}$(GeV)}
  \psfrag{M}{$M_{1}$(GeV)}
  \psfrag{-}{-}
  \psfrag{0}{0}
  \psfrag{1}{1}
  \psfrag{2}{2}
  \psfrag{3}{3}
  \psfrag{4}{4}
  \psfrag{5}{5} 
  \psfrag{6}{6}
  \psfrag{7}{7}
  \psfrag{8}{8}
  \psfrag{9}{9}
  \psfrag{mn1mch1sugramu200}{}
  \subfigure[$\tan\beta=10$]{\includegraphics[height=5.0cm]{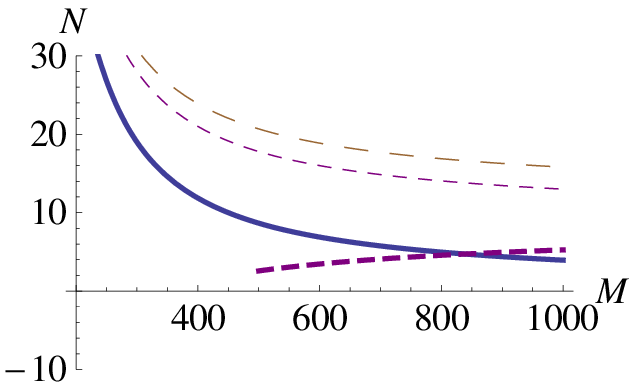}}
  \subfigure[$\tan\beta=30$]{\includegraphics[height=5.0cm]{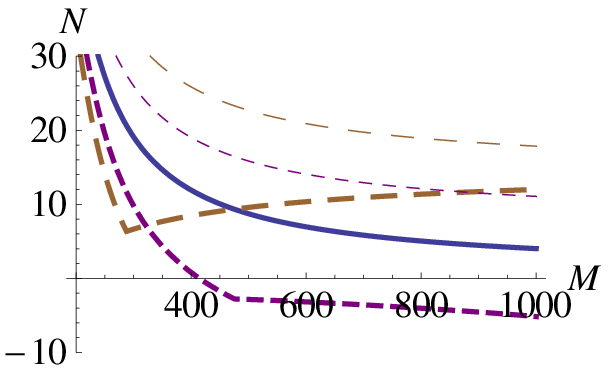}}
\caption{The difference between the lightest neutralino and the lightest 
chargino mass in mSUGRA plotted for several values of 
the parameter $\epsilon_1 = \frac{\lambda}{M}\mu$.  The blue solid line corresponds to 
$\epsilon_1 = 0$, and the thick dashed lines in  order of increasing 
dash length represent $\epsilon_1 = 0.05$ (violet), $\epsilon_1 = 0.1$ (ochre). 
The thin dashed lines correspond to  $\epsilon_1 = -0.05$ (violet), $\epsilon_1 = -0.1$ 
(ochre), again in the order of increasing dash length. Here $\mu=200$ GeV and one-loop
radiative corrections are included.}
\label{fig:mn1mch1}
\end{figure}

In Fig.~\ref{fig:mn1} we show the lightest neutralino and chargino masses 
for several values of
$\epsilon_1$,  $\epsilon_1=0,\, \pm 0.05,\, \pm0.1$.
We have  plotted these masses for the mSUGRA model.  
The dimension five operator causes a shift in the lightest Higgs mass which
can bring it down below the current experimental limit 
\cite{Berg:2009mq}. 
We have excluded the parts of the graphs where $m_{h^0}<111$ GeV in Figs. 2-6, when calculating the Higgs mass with SOFTSUSY~\cite{SOFTSUSY} and shift caused by dim 5 operators is taken into account.
Because for $\mu<<M_1,M_2$ the higgsino sector strongly dominates the 
lightest neutralino and chargino masses, and thus the plot  for mSUGRA is a representative for the mirage mediation
models as well since the only difference in the masses in these models  is due to the gaugino nonuniversality.
It is seen that the effect of BMSSM operators in the case of 
mSUGRA pattern of gaugino masses is a few GeV, depending on the parameters.
For $\tan\beta =10$, Figs.~2~ (a) and (c), for positive $\epsilon_1=0.05$, there are experimentally
allowed Higgs masses only 
for $M_1> 450$. 
For $\epsilon_1=0.1$
Higgs is too light for all $M_1\leq 1$ TeV.
Increasing $\tan\beta$ leads to a heavier Higgs, and the $M_1$ values shown in 
Figs.~2~(b) and (d) are allowed.
The effect of dimension five operator for small $M_1$ values is opposite for neutralino and 
chargino masses, while for large 
values of $M_1$, the neutralino mass is always smaller than what it is without the  $\epsilon_1$  correction.
For chargino mass the correction is  positive for negative $\epsilon_1$ and it
is negative for
positive $\epsilon_1$.
Thus, the effect of dimension five operator is enhanced for 
negative $\epsilon_1$ in the difference of chargino and neutralino masses,
as seen in Fig.~\ref{fig:mn1mch1}.
We have not shown the results for the AMSB case, since
in the AMSB  $\mu$ cannot be smaller than $M_1$ due to the electroweak 
symmetry breaking condition~\cite{Gherghetta:1999sw},
and thus in this case the dimension five contribution is negligible to the lightest
neutralino and chargino masses. 

If the $\mu$ parameter is large compared to the soft gaugino masses, 
the two heaviest of the neutralinos are mostly higgsinos.
The relative 
contribution of the dimension five operator to the mass for a heavy
particle from the BMSSM operators is small.
We conclude that if dimension 5 contribution to the masses of neutralinos 
and charginos is sizable, one cannot use purely the neutralino and chargino masses 
to determine the supersymmetry breaking mechanism.
We, therefore, consider here two different sum rules involving neutralino
and chargino masses and their squares. The dependence on gaugino masses
enters these sum rules in a specific manner. 

From the trace of the neutralino mass matrix (\ref{neutmatrix}) one 
obtains the sum over the neutralino mass eigenvalues which we denote 
by $\sigma$. This can be written as 
\bea
\sigma(\epsilon_1) \equiv \sum_{i = 1}^4 \eta_i m_{\tilde\chi^0_i}
=M_1+M_2+2 \frac{\epsilon_1}{\mu} v^2,
\label{sum1}
\eea
at  leading order in  $\epsilon_1$, where  $\eta_i$ is the sign of 
the $i$th eigenvalue.
This sum rule depends on the $\mu $ parameter through BMSSM 
operators, when $\epsilon_1$ is taken as an independent parameter.
It should be noted that in most of the allowed parameter space
the neutralino mass matrix has one negative eigenvalue 
(see Table~\ref{negativeneu} for the gluino masses we use in this work). 
This needs to be taken into account when evaluating the sum.
An advantage of this sum rule is that in addition to the gaugino mass
parameters and $\epsilon_1$, it depends only on the supersymmetric higgsino mixing
parameter $\mu$.
\begin{center}
\begin{table}[h]
\parbox{.45\linewidth}{
\centering
\begin{tabular}{|l|c|r|}
\hline
${M_{\tilde g}}$(GeV) & 750 & 2000 \\
\hline\hline
mSUGRA & $\Neu{3}$ & $\Neu{2}$ \\ \hline
AMSB & $\Neu{2}$ & $\Neu{2}$ \\ \hline
mirage $\alpha=1$ &  $\Neu{2}$ &  $\Neu{1}$ \\ \hline
mirage $\alpha=2$ &  $\Neu{1}$ &  $\Neu{1}$ \\ 
\hline
\end{tabular}
\caption{The eigenvalue of the neutralino mass matrix with a negative sign.}
}
\label{negativeneu}
\hfill
\parbox{.45\linewidth}{
\centering
%
\begin{tabular}{|l|r|r|r|r|}
\hline
\multicolumn{1}{|l|}{${M_{\tilde g}}$(GeV)} & \multicolumn{2}{c}{750} & \multicolumn{2}{|c|}{2000}\\ \hline
$\tan\beta$ & 10 & 30 & 10 & 30\\ \hline\hline
mSUGRA & $-0.06$ & $<-0.1$ & $0.03$ & $-0.06$\\ \hline
AMSB & $-0.04$ & $-0.09$  & $-0.02$ & $-0.04$\\ \hline
mirage $\alpha=1$ & $0.02$ & $0.07$ & $0.00$ & $0.00$\\ \hline
mirage $\alpha=2$ & $-0.04$ & $<-0.1$ & $-0.01$ & $-0.03$\\ 
\hline
\end{tabular}
\caption{The value of $\epsilon_1$ corrensponding to Higgs mass of 125 GeV. Higgs mass increases with
decreasing $\epsilon_1$. }
}
\label{higgs125neu}
\end{table}
\end{center}
Using relations (\ref{gaugino2}), 
(\ref{gmass}), (\ref{miragelowe}), and (\ref{mirageuni} ) the gaugino 
mass parameters $M_1$ and $M_2$ can be expressed in terms of the gluino mass 
${M_{\tilde g}}$ and coupling constant $\alpha_i$, both observable quantities.
For mSUGRA, AMSB and mirage mediation the sum rule can then be written as,
with $B=\ln(M_{GUT}/M_{mir})/(16\pi^2)$,
\bea
\sigma_{mSUGRA}(\epsilon_1)
&=&\frac{M_{\tilde g}}{\alpha_3}(\alpha_1 + \alpha_2) + 2 \frac{\epsilon_1}{\mu} v^2,\nonumber \\
\sigma_{AMSB}(\epsilon_1)
&=&\frac{M_{\tilde g}}{3}\left[\frac{\alpha_2}{\alpha_3}
+\frac{33}{5}\frac{\alpha_1}{\alpha_3}\right] + 2\frac{\epsilon_1}{\mu} v^2,\nonumber \\
\sigma_{mirage}(\epsilon_1)
&=&\frac{M_{\tilde g}}{\alpha_3}\left[{1-3B}\right]^{-1}\left[\alpha_2\left(1+B\right)\right.
+\alpha_1\left.\left(1+\frac{33}{5}B\right)\right]
+  2 \frac{\epsilon_1}{\mu} v^2.
\label{smallsum}
\eea

\begin{figure}
\psfrag{e}{$\epsilon_1$}
\psfrag{1}{1}
\psfrag{2}{2}
\psfrag{3}{3}
\psfrag{4}{4}
\psfrag{5}{5}
\psfrag{0}{0}
\psfrag{-}{-}
\psfrag{.}{.}
\psfrag{S}[][][1.1]{$\frac{\sigma(\epsilon_1)-\sigma(0)}{\sigma(\epsilon_1)}$}
\subfigure[ $M_{\tilde g}=750$; $\mu=200$ GeV.]{
\includegraphics[height=5.0cm]{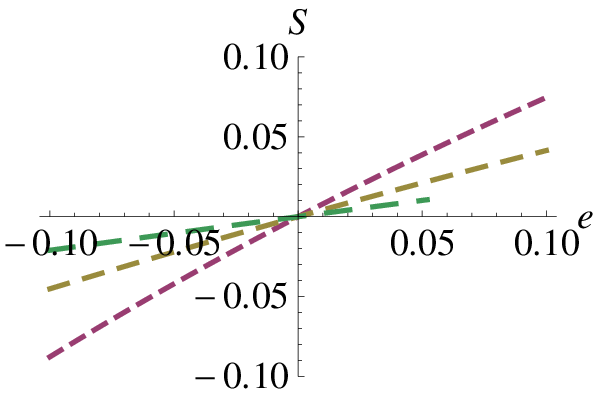}}
\subfigure[ $M_{\tilde g}=750$; $\mu=500$ GeV.]{
\includegraphics[height=5.0cm]{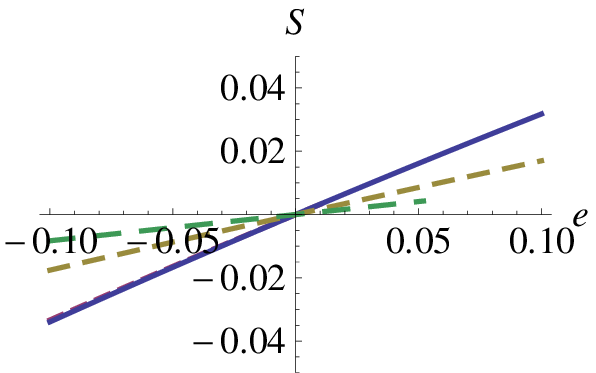}}
\subfigure[ $M_{\tilde g}=2000$; $\mu=200$ GeV.]{
\includegraphics[height=5.0cm]{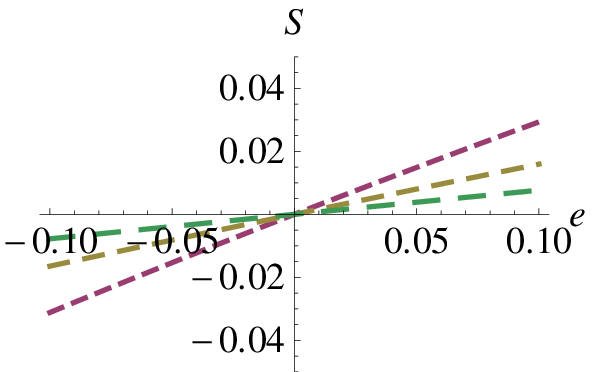}}
\subfigure[ $M_{\tilde g}=2000$; $\mu=500$ GeV.]{
\includegraphics[height=5.0cm]{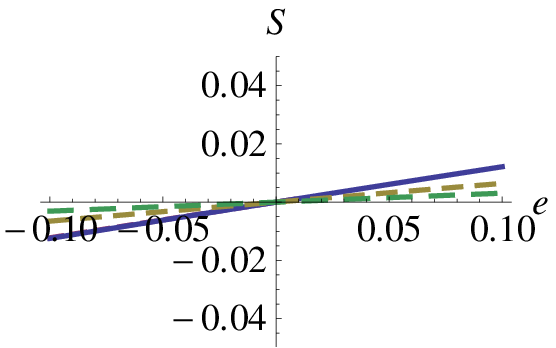}}

\caption{The contribution arising from 
$\epsilon_1$ to the total sum of (\ref{sum1}) in different supersymmetry breaking models. The solid blue line corresponds to AMSB;
mSUGRA~(violet), and mirage mediation with $\alpha=1$~(ochre), and $\alpha=2$~(green) models, respectively, are presented in the order of increasing dash length. 
}
\label{sum1plot}
\end{figure}

In Fig.~\ref{sum1plot} we have plotted the magnitude of the dimension five contribution relative to the whole sum 
with two $\mu$ and $M_{\tilde g}$ values, $\mu= 200, 500$ GeV, and  $M_{\tilde g}= 750, 2000$ GeV.
The plotted quantities can be written in terms of observables as 
\bea
\frac{\sigma(\epsilon_1)-\sigma(0)}{\sigma(\epsilon_1)}&=&\frac{\sum_{i=1}^4 \eta_im_{\Neu{i}}-\gamma_{SB} M_{\tilde g}}{\sum_{i} \eta_im_{\Neu{i}}},
\label{dim5per1}
\eea
where $\gamma_{SB}$ refers to the coefficient of $M_{\tilde g}$ in different gaugino mass patterns in Eq.~(\ref{smallsum}).
We have again taken account of the experimental limit for the Higgs mass by excluding the parts of the lines violating the limit
of $m_h<111$ GeV (when calculating $m_h$, we use $\tan\beta=30$).
AMSB is not allowed for the $\mu=200$ GeV case due to the constraint $\mu>M_1$ in this model.
In the sum $\sigma$ the dimension five contribution is inversely proportional to $\mu$, and the maximum
percentage contribution is achieved with the lowest gluino mass.
The contribution is largest for mSUGRA pattern, and smallest for 
mirage mediation with $\alpha=2$.
In our example with $M_{\tilde g}=750$ GeV and $\mu=200$ GeV, the contribution with $\epsilon_1=-0.1$ varies between
-2.5 \% and -9 \% . 

The Higgs mass is an important constraint for the 
breaking patterns that we have studied in this paper.
For the chosen values of $\tan\beta=10, 30$, and gluino masses $m_{\tilde{g}}=750$ GeV and 2 TeV, we have shown in Table~\ref{higgs125neu} 
the values of $\epsilon_1$ for which $m_h=125$ GeV.
The smaller the  $\epsilon_1$ parameter is, the heavier the Higgs is.
For mSUGRA and mirage mediation with $\alpha=2$ and for $\tan\beta=30$, $m_{\tilde{g}}=750$ GeV, the required $\epsilon_1$
would be smaller than -0.1.

From the trace of the squares of the neutralino and chargino
mass matrices, one obtains a sum rule for the neutralino and chargino masses squared, which we denote by $\Sigma$:
\bea
\Sigma(\epsilon_1)
\equiv 2\sum_{i = 1}^2 m_{\tilde \chi^\pm_i}^2 
- \sum_{i = 1}^4 m_{\tilde\chi^0_i}^2
& = & \left[M_2^2 - M_1^2 \right] + 4M_W^2 - 2M_Z^2
+ 4 \epsilon_1 v^2 \sin 2 \beta.
\label{sum2}
\eea
at  leading order in  $\epsilon_1$. 
This sum rule depends on $\tan\beta$ in addition to $M_1$, $M_2$ and $\epsilon_1$ but not on $\mu$.
In this sense the sum rules (\ref{sum1}) and (\ref{sum2}) are complementary.

The dimension 5 contribution in $\Sigma(\epsilon_1)$ decreases for increasing $\tan\beta$.
The gaugino 
mass parameters $M_1$ and $M_2$ can again be expressed 
in terms of the gluino mass 
${M_{\tilde g}}$ and coupling constants $\alpha_i$.
For mSUGRA, AMSB and mirage mediation the sum rule  can be written as

\bea
\Sigma_{mSUGRA}(\epsilon_1)
&=&\frac{M_{\tilde g}^2}{\alpha_3^2}(\alpha_2^2 - \alpha_1^2) 
+ 4M_W^2 - 2M_Z^2 + 4 \epsilon_1 v^2 \sin 2 \beta,\nonumber \\
\Sigma_{AMSB}(\epsilon_1)
&=&\frac{M_{\tilde g}^2}{9}\left[\frac{\alpha_2^2}{\alpha_3^2}
-(\frac{33}{5})^2\frac{\alpha_1^2}{\alpha_3^2}\right] + 4M_W^2 - 2M_Z^2 
+ 4 \epsilon_1 v^2 \sin 2 \beta,\nonumber \\
\Sigma_{mirage}(\epsilon_1)
&=&\frac{M_{\tilde g}^2}{\alpha_3^2}\left[1-3B\right]^{-2}\left[\alpha_2^2\left(1+B\right)^2
-\alpha_1^2
\left(1+\frac{33}{5}B\right)^2\right]\nonumber\\
&&+ 4M_W^2 - 2M_Z^2+ 4 \epsilon_1 v^2 \sin 2 \beta.
\label{largesum}
\eea

\begin{figure}
\psfrag{e}{$\epsilon_1$}
\psfrag{.}{.}
\psfrag{0}{0}
\psfrag{1}{1}
\psfrag{2}{2}
\psfrag{3}{3}
\psfrag{4}{4}
\psfrag{5}{5}
\psfrag{-}{-}
\psfrag{S}[][][1.1]{$\frac{\sum(\epsilon_1)-\sum(0)}{\sum(\epsilon_1)}$}
\psfrag{MSUGRA}{mSUGRA}
\psfrag{AMSB}{AMSB}
\psfrag{MIRAGEa1}{Mirage, $\alpha=1$}
\psfrag{MIRAGEa2}{Mirage, $\alpha=2$}
\subfigure[ $M_{\tilde g}=750$ GeV; $\tan \beta = 10$. ]{
\includegraphics[height=5.0cm]{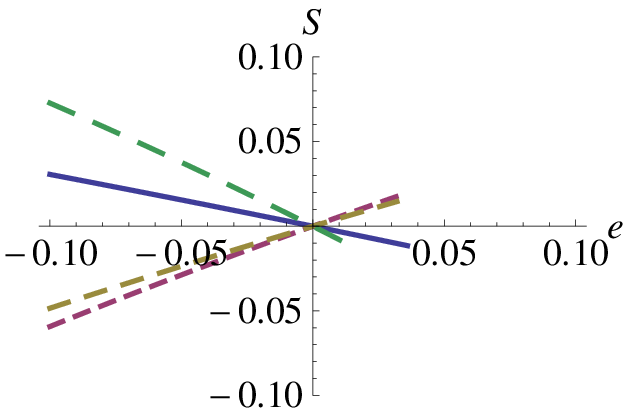}}
\subfigure[ $M_{\tilde g}=750$ GeV; $\tan \beta = 30$.]{
\includegraphics[height=5.0cm]{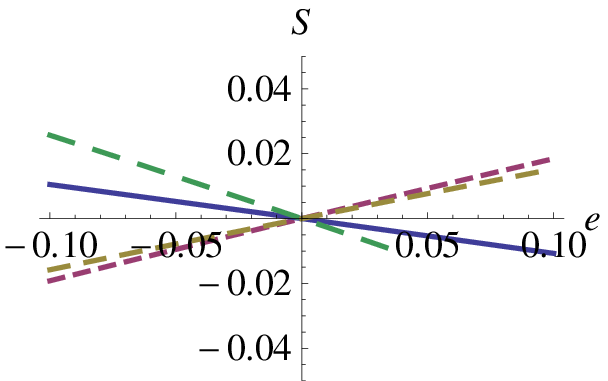}}
\subfigure[ $M_{\tilde g}=2000$ GeV; $\tan \beta = 10$.]{
\includegraphics[height=5.0cm]{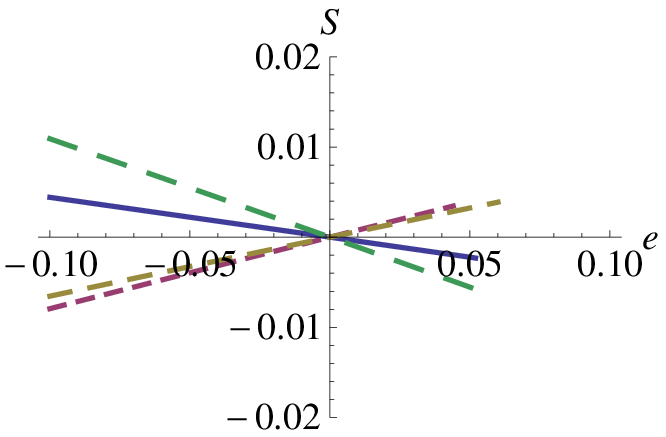}}
\subfigure[ $M_{\tilde g}=2000$ GeV; $\tan \beta = 30$.]{
\includegraphics[height=5.0cm]{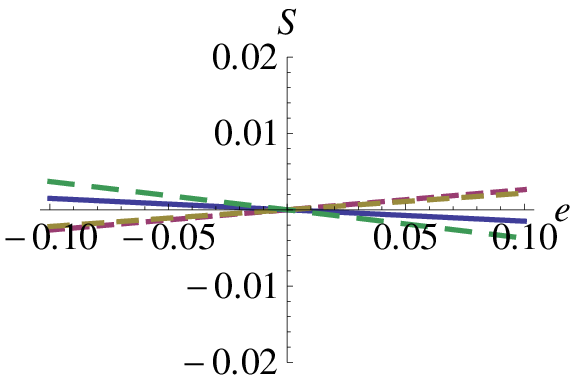}}

\caption{The contribution arising from 
$\epsilon_1$ to the total sum of (\ref{sum2}) in different supersymmetry breaking models. The solid blue line corresponds to AMSB;
mSUGRA~(violet), and mirage mediation with $\alpha=1$~(ochre), and $\alpha=2$~(green) models, respectively, are presented in the order of increasing dash length.}
\label{sum2plot}
\end{figure}

In Fig.~\ref{sum2plot} we have plotted the magnitude of the dimension five contribution relative to the whole sum 
with two $\tan\beta$ and $M_{\tilde g}$ values, $\tan\beta=10, 30$ and  $M_{\tilde g}= 750, 2000$ GeV.  
The plotted quantities can be written in terms of observables as 
\bea
\frac{\Sigma(\epsilon_1)-\Sigma(0)}{\Sigma(\epsilon_1)}&=&2\frac{\sum_{i = 1}^2 m_{\tilde \chi^\pm_i}^2 
- \sum_{i = 1}^4 m_{\tilde\chi^0_i}^2-\alpha_{SB}^2 M_{\tilde g}^2}{\sum_{i = 1}^2 m_{\tilde \chi^\pm_i}^2 
- \sum_{i = 1}^4 m_{\tilde\chi^0_i}^2},
\label{dim5per2}
\eea
where $\alpha_{SB}$ is the supersymmetry breaking model dependent coefficient of $M_{\tilde g}^2$ in (\ref{largesum}).
As seen from Fig. \ref{sum2plot} increasing $\tan\beta$ from 10 to 30
rougly halves the dimension five contribution. 
Larger $\tan\beta$ however allows larger positive values $\epsilon_1$ without violating the Higgs mass constraint. In contrast with $\sigma$, the 
maximum dimension five contribution of 10 \% is seen in the mirage mediation model with $\alpha=2$, and in mSUGRA the contribution is the lowest
of the four examined models.
It is seen that for AMSB and mirage mediation with $\alpha=2$ the contribution to $\sigma$ is opposite sign 
to the contribution to $\Sigma$, while for mSUGRA and mirage mediation with $\alpha=1$, $\sigma$ and $\Sigma$ have the same sign.

By combining the sum rules Eq.~(\ref{smallsum}) and (\ref{largesum}) we obtain a relation for $\tan \beta$ and $\mu$ that is independent of $\epsilon_1$,
\bea
\mu=\frac{2\sum_{i = 1}^2 m_{\tilde \chi^\pm_i}^2 
- \sum_{i = 1}^4 m_{\tilde\chi^0_i}^2-\alpha_{SB}^2 M_{\tilde g}^2 - 4M_W^2 + 2M_Z^2}{\sum_{i=1}^4 \eta_i m_{\Neu{i}}-\gamma_{SB} M_{\tilde g}}\frac{1+\tan^2\beta}{4\tan\beta}. \label{linear}
\eea
This relation can be used for estimating the value of $\mu$ in BMSSM models 
if $\tan\beta$ is known.
It should be noted that this formula does not exist without the BMSSM operator $\epsilon_1$.
Thus a consistent value with other measurements may indicate 
the existence of the BMSSM operators.
From precise measurements the value of $\epsilon_1$ can also be determined from Eq.~(\ref{smallsum}) 
and (\ref{largesum}) when $\mu$ or $\tan\beta$ are known.

\begin{figure}
\psfrag{S}[][][1.2]{$\frac{\sum(\epsilon_1)-\sum(0)}{\sum(\epsilon_1)}$}
\psfrag{0}{0}
\psfrag{1}{1}
\psfrag{5}{5}
\psfrag{2}{2}
\psfrag{.}{.}
\psfrag{3}{3}

\psfrag{-}{-}
\psfrag{F}{mSUGRA}
\psfrag{A}{AMSB}
\psfrag{E}{$\alpha=1$}
\psfrag{G}{$\alpha=2$}
\psfrag{C}{$M_1=200$ GeV}
\psfrag{B}{$M_1=100$ GeV}
\psfrag{D}{$M_1=400$ GeV}
\psfrag{EPSI2A}{$\epsilon_1=-0.05$}
\psfrag{M}{$M_{2}/M_{1}$}
\includegraphics[height=7cm]{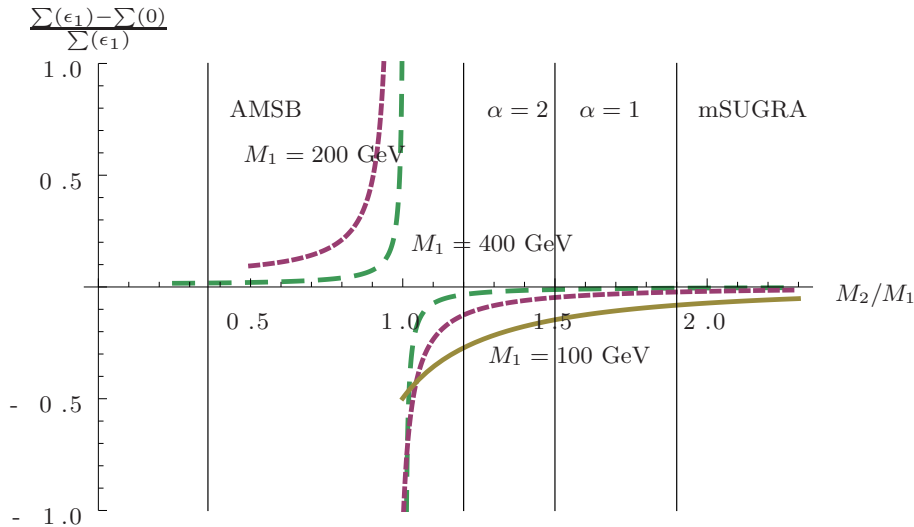}
\caption{The fraction of the contribution arising from 
$\epsilon_1$ to the total
sum of (\ref{sum2}) plotted as function of the ratio  
$M_2/M_1$ with  $M_1=400$ GeV (long dashes, green), $M_1=200$ GeV (short dashes, purple), 
and $M_1=100$ GeV (solid line, ochre). On the horizontal 
axis $M_{2}/M_{1}=0.36$ corresponds to AMSB,
$M_{2}/M_{1}=1.2$ to mirage mediation with $\alpha=2$, $M_{2}/M_{1} = 1.5$ 
to mirage mediation with $\alpha = 1$, and $M_{2}/M_{1} = 1.9$ to mSUGRA. 
Here $\epsilon_1=-0.1$ and $\tan \beta =10$. Only the parts of the lines that agree with the experimental
limit for the chargino mass~(\ref{ch-limit}) are shown.}
\label{sum2m2}
\end{figure}

The gaugino mass pattern realized in Nature may well turn out to be a mixture of the patterns
studied here.
This possibility can be considered by a general study of the ratio of $M_1$ and $M_2$.
In Fig.~\ref{sum2m2} we show the fraction of the contribution from 
the dimension five operator 
to the sum rule~(\ref{sum2}) for $\epsilon_1=-0.1$ as a function of the ratio of the mass 
parameters $M_2$ and $M_1$. Although at $M_1=400$ GeV (and larger)
the dimension five contribution remains at less than a few percent for all models, $M_1=100$ GeV can produce as high
as a 20 percent dimension five contribution in mirage mediation with $\alpha=2$ and a 10 percent contribution in mSUGRA.
As expected, the contribution is highest near the point $M_2/M_1=1$, where 
the sum of the squares of the gaugino mass parameters cancels in the sum rule, thus making the 
sum completely independent of the gaugino masses. This point corresponds to 
mirage mediation with $\alpha=2.17$. 
Consequently, mirage mediation models with $\alpha$ close to this value allow significant dimension five 
contributions, although the lower bound for the gluino mass restricts $M_1$ to 1 TeV range and above. 
The experimental limit for the chargino mass rules out $M_1$ lower than 280 GeV in AMSB, and the dimension five
contribution remains at a few percent for all allowed values for the gaugino masses for this model.

\begin{figure}
\psfrag{X}{$M_{\tilde g}$(GeV)}
\psfrag{x}{$M_{\tilde g}$(GeV)}
\psfrag{M}{$M_{\tilde g}$(GeV)}
\psfrag{m}{$M_{\tilde g}$(GeV)}
\psfrag{q}{$\frac{\Delta\sigma(0)}{\sigma(0)}$}
\psfrag{Q}{$\frac{\Delta\sum(0)}{\sum(0)}$}
\psfrag{0}{0}
\psfrag{1}{1}
\psfrag{2}{2}
\psfrag{3}{3}
\psfrag{4}{4}
\psfrag{5}{5}
\psfrag{6}{6}
\psfrag{7}{7}
\psfrag{8}{8}
\psfrag{9}{9}
\psfrag{-}{-}
\psfrag{Y}[][][1.1]{$\sum(0)$(GeV$^2$)}
\psfrag{y}{$\sigma(0)$(GeV)}
\subfigure[$\tan \beta = 10$, $\mu = 500$ GeV]{
\includegraphics[height=4.6cm]{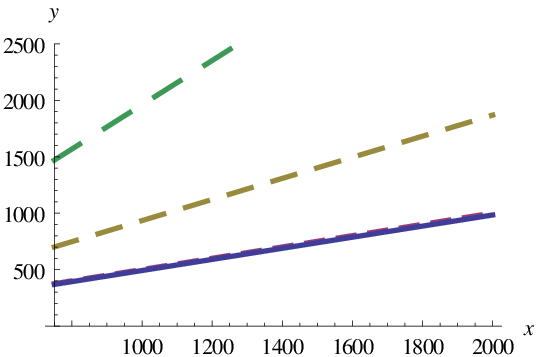}}
\subfigure[$\tan \beta = 10$, $\mu = 500$ GeV]{
\includegraphics[height=4.6cm]{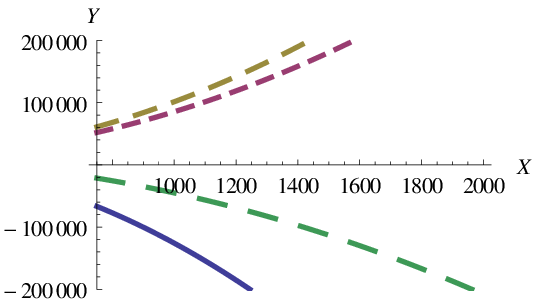}}
\subfigure[$\tan \beta = 10$, $\mu = 500$ GeV]{
\includegraphics[height=5.0cm]{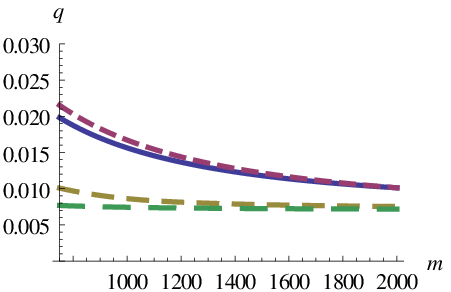}}
\subfigure[$\tan \beta = 10$, $\mu = 500$ GeV]{
\includegraphics[height=5.0cm]{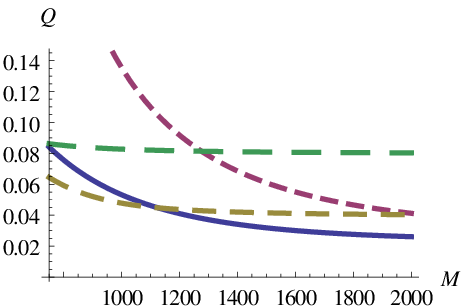}}
\subfigure[$\tan \beta = 10$, $\mu = 200$ GeV]{
\includegraphics[height=5.0cm]{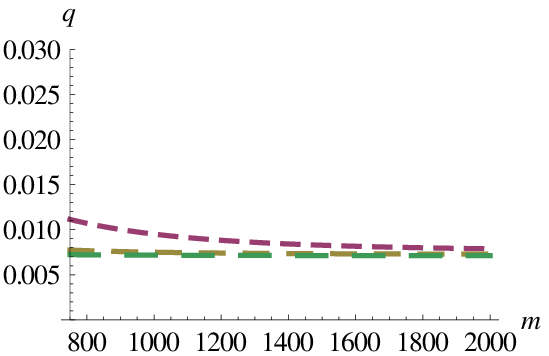}}
\subfigure[$\tan \beta = 30$, $\mu = 500$ GeV]{
\includegraphics[height=5.0cm]{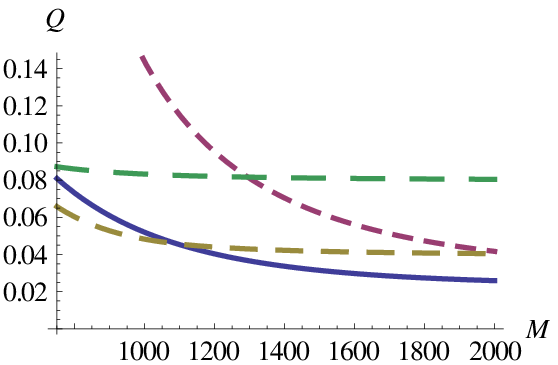}}
\caption{The quantities \ref{smallsum} and \ref{largesum} as a function of the gluino mass and their experimental uncertainties assuming 1\% uncertainty in the measurement
of three heaviest neutralino masses and of both chargino masses.
The solid blue line corresponds to AMSB; mSUGRA~(violet), and mirage mediation with $\alpha=1$~(ochre), and $\alpha=2$~(green) models, 
respectively, are presented in the order of increasing dash length.}
\label{sumabs}
\end{figure}

The usefulness of the sum rules depends on the accuracy with which the masses can be measured. 
The experimental error in the measurement of
the neutralino and the chargino masses has been discussed in e.g.~\cite{rainwater} for the LHC
and for a possible future linear collider. 
While the quoted accuracies are not precise enough for using the sum rules, we have calculated as 
an example
the accuracy for \ref{smallsum} and \ref{largesum} assuming 1 \% error in the measurement of the 
three heaviest neutralino masses and 
in both chargino masses, while neglecting the error in the lightest neutralino mass. 
Results 
are presented in Fig. \ref{sumabs}. The accuracy of measuring $\Sigma$ is diminished by the negative contribution of
the neutralinos in the sum as well as the squaring of the masses,
although at low gluino masses the uncertainty is of the same order of magnitude as the maximum $\epsilon_1$ contribution in our range of $\epsilon_1>-0.1$ in AMSB and mirage mediation models.

The accuracy of $\sigma$ is affected by the mass of the neutralino with negative contribution to the sum compared to the masses of the other three neutralinos. We note that the uncertainty in $\sigma$
differs significantly with respect to the $\mu$ parameter only in the case of mSUGRA, and is largely independent of the gluino mass for $\mu=200$ GeV.  Since the $\epsilon_1$ contribution is inversely proportional 
to $\mu$, the usefulness of $\sigma$ in the detection of any BMSSM effect is greater for lower values of $\mu$, for which the uncertainty is at 1\% level for the whole gluino mass range (and in
all models, excluding AMSB). As a comparison, the BMSSM contribution ranges from 1 \% to 4 \% for $\epsilon_1=-0.05$, and from 2 \% to 9 \% for $\epsilon_1=-0.1$, when $M_{\tilde g}=750$ GeV and $\mu=200$ GeV (Fig.\ref{sum1plot}).

\section{Summary}

We have studied the contribution of  the dimension five
BMSSM operators involving chiral Higgs superfields to  the neutralino 
and chargino masses.
The contribution can be significant when the higgsino mixing parameter $\mu$
is small compared
to the soft supersymmetry breaking gaugino mass parameters, as we have 
illustrated.
If the $\mu$ parameter is large, its effect is negligible on the mass
of the lightest neutralino, which is dominantly a gaugino.
Thus, the sensitivity to the BMSSM operator studied here 
is very different in different supersymmetry breaking models, 
since in the mSUGRA and mirage mediation models the $\mu$ 
parameter can be small, while in the anomaly mediation models 
it is always larger than the gaugino mass parameters.
The effect of the dimension five operators on the masses of
the heavier neutralinos is relatively small as compared to the
lightest neutralino  mass, and thus more difficult to isolate.

We have examined whether the sum rule involving squares of the neutralino and chargino masses
and the sum rule involving neutralino masses
could be used for the detection of BMSSM operators 
by calculating the contribution of the dimension
five parameter to the sums.
We have shown that the two sum rules can be combined to derive a relation between $\mu$ and $\tan\beta$ 
which is valid in the presence of the studied dimension 5 BMSSM operator.

The accuracy of the neutralino and chargino mass measurements is a key issue in the usefulness of the 
sum rules.
We have examined whether the sum rule  ($\Sigma$) involving squares of the neutralino and chargino masses
and the sumrule ($\sigma$) involving neutralino masses could be used for the detection of BMSSM by calculating the contribution of the dimension
five parameter to the sum, and evaluating the accuracy to which the sum can be
measured using the anticipated accuracies for neutralino and chargino measurements at a linear collider. 
For large $\mu$ the
BMSSM effect contributes to the $\Sigma$-sum more significantly than to the lightest neutralino mass, but the cumulative error from
the squares of the neutralino and chargino masses diminishes the accuracy of
the total sum measurement. The uncertainty is at best of the same order of magnitude with the BMSSM contribution.
The other sum rule $\sigma$
involving neutralino masses has the advantage of having far less experimental uncertainty,
and for our example accuracies, the measurement error would be smaller than the dimension five contribution to the sum rule.

\bigskip
{\bf Acknowledgments}
KH and PT gratefully acknowledge support from the Academy of Finland 
(Project No. 137960).
PNP would like to thank Department of Physics, University of Helsinki, and 
Helsinki Institute of Physics for hospitality
while part of this work was done. The work of PNP is supported by
the J. C. Bose National Fellowship, and by the Council of Scientific and
Industrial Research, India under the project No 03(1220)/12/EMR-II.

\end{document}